\documentclass[aip,ltp,reprint]{revtex4-1}
\usepackage{graphicx}
\draft

\begin{document}

\title{Stochastic amplification of weak signals in an RF SQUID with ScS contact}

\author{O.G. Turutanov}
\email{turutanov@ilt.kharkov.ua}
\affiliation{B. Verkin Institute for Low Temperature Physics and Engineering, NAS of Ukraine, Lenin Ave. 47, Kharkov 61103, Ukraine} %
\author{V.Yu. Lyakhno}
\affiliation{B. Verkin Institute for Low Temperature Physics and Engineering, NAS of Ukraine, Lenin Ave. 47, Kharkov 61103, Ukraine} %
\author{V.I. Shnyrkov}
\affiliation{B. Verkin Institute for Low Temperature Physics and Engineering, NAS of Ukraine, Lenin Ave. 47, Kharkov 61103, Ukraine} %

\date{\today}

\begin{abstract}
The paper presents results of a numerical simulation of the
stochastic dynamics of magnetic flux in an RF SQUID loop with a
Josephson point (ScS) contact driven by a mix of band-limited
Gaussian noise and low-frequency small-amplitude sine signal, at
finite temperatures  $0<T<T_{c}$. A change in the gain of the
stochastically amplified weak signal with the temperature rise is
shown to be due to smearing of the energy barrier between adjacent
metastable current states of the loop. A comparison to an RF SQUID
with a tunnel (SIS) junction is done.
\end{abstract}



\maketitle

\section{Introduction}

The magnetometers based on Superconducting Quantum Interference
Devices (SQUIDs) are widely used in physical experiments, medicine
(magnetocardiographs and magnetoencephalographs), geophysics
(underground radars), electronic manufacturing (SQUID microscopes).
The sensitivity of SQUIDs and their quantum analogues, SQUBIDs, has
practically reached the quantum limitation
\cite{1Ketchen,2Shnyrkov,3Shnyrkov}. However, with increase of the
quantizing loop inductance up to  $L\sim 10^{-9} -10^{-10} \; H$ ,
the thermodynamic fluctuations lead to expressed degradation of the
energy resolution.

As shown earlier \cite{4Rouse,5Hibbs,6Turutanov,7Glukhov}, the
sensitivity of the magnetometers can be considerably enhanced in
this case by using the stochastic resonance (SR) effect. The SR
phenomenon discovered in the early 1980s \cite{8Benzi,9Nicolis}
manifests itself in a non-monotonic rise of a nonlinear (often
bi-stable) system response to a weak periodic signal which peaks at
a certain intensity of the noise added (or inherent) to the system.

Owing to extensive studies during the last two decades, the
stochastic resonance effect has been revealed in a variety of
natural and artificial systems, both classical and quantum ones.
Analytical approaches and quantifying criteria for estimation of the
noise-induced ordering were determined and then summarized in the
reviews \cite{10Gammaitoni,11Anishchenko}. For example, the gain of
40 dB was experimentally demonstrated \cite{4Rouse} for a weak
harmonic signal due to SR in a SQUID with the tunnel (SIS) Josephson
junction. For an under-optimal noise intensity, the signal gain in
SR SQUIDs can be nevertheless maximized with the
stochastic-parametric resonance (SPR) effect \cite{12Turutanov}
emerging when the system is affected by the weak signal, the noise
and an additional high-frequency electromagnetic field
simultaneously.

Not so long ago, researchers paid their attention to clean
small-sized superconducting contacts of "constriction" (ScS) type
called the "atomic-size contacts", ASCs \cite{13Agrait}. They have
only a few quantum conductive channels in their cross section so
they often referred to as the "quantum point contacts", QPCs
\cite{14Beenakker}. The critical current  $I_{c} $  of such contacts
may take discrete values. The interest to the QPCs is roused by
studies of the channel quantum conductivity and building
superconducting qubits with large energy level splitting  $\Delta
E/h\sim 30\; GHz$ \cite{2Shnyrkov}. The current-phase relation (the
supercurrent $I_{s} $  as a function of the order parameter phase
difference $\varphi $ ) in the clean ScS contacts with ballistic
electron transit at low temperatures ( $T\to 0$ )
\cite{14Beenakker,15Kulik} substantially differs from the classical
Josephson current-phase relation for a SIS junction $I_{s}^{SIS}
(\varphi )=I_{c} \sin \varphi $  and features a "saw-like" shape
with discontinuities at $\varphi =n\pi$.

The coupling energies for SIS and ScS obtained by integrating the
current-phase relations therefore differ, too. If the SIS junction
is incorporated into a superconducting loop and an external magnetic
flux  $\Phi _{e} =\Phi _{0} /2$  is applied ( $\Phi _{0}
=h/2e\approx 2.07\cdot 10^{-15} \; Wb$  is the magnetic flux
quantum), its current-phase relation forms the symmetric two-well
potential of the loop, which is needed for SR, only at sufficiently
large loop inductances and/or the junction critical current, namely,
when  $\beta _{L} =2\pi LI_{c} /\Phi _{0} >1$ . Unlikely, the
potential energy  $U^{ScS} (\Phi )$  of the superconducting loop
with a QPC has a "sharp" barrier with the singularity in its top
separated two metastable current-carrying states of the loop with
different intrinsic magnetic flux  $\Phi $  for any  $\beta _{L} $
including  $\beta _{L}<<1$.

In classical case for the zero-temperature limit, the stochastic
dynamics of the magnetic flux in a superconducting loop with the
Josephson ScS contact (ScS SQUID) substantially differs from that
for SIS SQUID studied earlier \cite{4Rouse,5Hibbs,7Glukhov}. Taking
into account the quantum fluctuations at finite temperatures
$0<T<T_{c}$ changes the current-phase relation for the ScS contact
and smoothes the potential barrier in the loop with ScS contact
\cite{17Khlus}. In this work we analyze the stochastic amplification
of weak low-frequency harmonic signals in a superconducting loop
interrupted by Josephson ScS contact, at finite temperatures and
various  $\beta _{L}$. The results are compared with the stochastic
dynamics of a SIS SQUID.

\section{Model}

The stochastic dynamics of the magnetic flux inside an RF SQUID loop
(inset in Fig.\,\ref{fig01}) is studied by numerical solving the
motion equation (Langevin equation) in the model of resistively
shunted junction (RSJ model):

\begin{figure}[h!]
\centering
\includegraphics[width = 0.8\columnwidth]{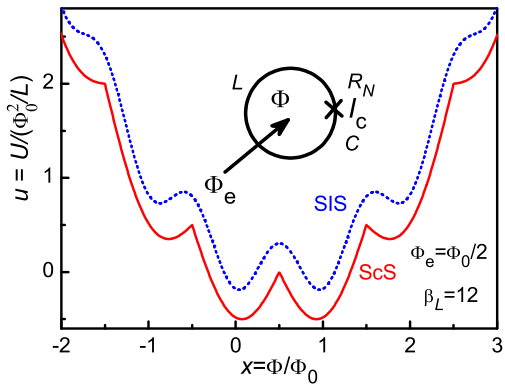}
\caption{\label{fig01}%
The potential energy of RF SQUID loop with ScS contact (solid line)
and SIS junction (dotted line) as a function of reduced magnetic
flux in the loop. The non-linearity parameter  $\beta _{L} =12$,
fixed external magnetic flux  $\Phi _{e} =\Phi _{0} /2$. The inset
shows schematically the RF SQUID loop with a Josephson junction, the
designations are described in the text.}
\end{figure}

\begin{equation}\label{1}
\displaystyle LC\frac{d^{2} \Phi (t)}{dt^{2} } +\frac{L}{R_{N} }
\frac{d\Phi (t)}{dt} +L\frac{dU(\Phi )}{d\Phi } =\Phi _{e} (t) ,
\end{equation}

\noindent
 where  $C$  is the capacitance,  $R_{N} $  and  $I_{c} $
are the normal shunt resistance and the critical current of the
Josephson junction, respectively,  $\Phi (t)$  is the magnetic flux
inside loop,  $U(\Phi ,\Phi _{e} )$  is the loop potential energy of
the which is the sum of the loop magnetic energy  $U_{M} $  and the
Josephson junction coupling energy  $U_{J}$. The time-dependent
external magnetic flux  $\Phi _{e} (t)$  applied to the loop
contains the fixed and the varying components including the noise
one. The junction coupling energy  $U_{J} $  is type-determined (for
SIS, ScS, SNS, etc.). The potential energy of a loop with the
"traditional" tunnel (SIS) junction is equal to \cite{18Barone}

\begin{equation}\label{2}
\displaystyle U^{SIS} (\Phi ,\Phi _{e} )=\frac{(\Phi -\Phi _{e}
)^{2} }{2L} -E_{J}^{SIS} \cos \frac{2\pi \Phi }{\Phi _{0} } ,
\end{equation}

\noindent
where  $E_{J}^{SIS} =I_{c} \Phi _{0} /2\pi $  is maximal
coupling energy of the tunnel Josephson junction.

We consider the dynamics of the magnetic flux in the RF SQUID loop
incorporating the clean ScS contact with ballistic electron transit
\cite{15Kulik}.

For classical \cite{15Kulik} and quantum \cite{14Beenakker} ScS
contacts with the critical current  $I_{c} $  the current-phase
relation in the zero-temperature limit  $T=0$

\begin{equation}\label{3}
\displaystyle I_{S}^{ScS} (\varphi )=I_{c} (\sin \frac{\varphi }{2})
\,sgn\,(\cos \frac{\varphi }{2} )
\end{equation}

\noindent has the "saw-like" shape with discontinuities at  $\varphi
=n\pi$. Correspondingly, the potential energy of the superconducting
loop with ScS contact is

\begin{equation}\label{4}
\displaystyle U^{ScS} (\Phi ,\Phi _{e} )=\frac{(\Phi -\Phi _{e}
)^{2} }{2L} -E_{J}^{ScS} \left|\cos \frac{\pi \Phi }{\Phi _{0} }
\right| ,
\end{equation}

\noindent
where  $E_{J}^{ScS} =I_{c} \Phi _{0} /\pi $  is the maximal
coupling energy of the Josephson ScS contact.  $U^{ScS} (\Phi ,\Phi
_{e} )$ has singularities in tops of the barriers separating the
adjacent local minima which correspond to the loop metastable
current states.

Introducing the dimensionless parameter of non-linearity

\begin{equation}\label{5}
\displaystyle \beta _{L} =2\pi LI_{c} /\Phi _{0} ,
\end{equation}

\noindent and reducing the fluxes by the flux quantum  $\Phi _{0}$:
$x=\Phi /\Phi _{0} $ ,  $x_{e} =\Phi _{e} /\Phi _{0}$, and the
potential energy by  $\Phi _{0}^{2} /2L$, expressions (\ref{2}) and
(\ref{4}) can be rewritten, correspondingly, as

\begin{equation}\label{6}
\displaystyle u^{SIS} (x,x_{e} )=\frac{(x-x_{e} )^{2} }{2}
-\frac{\beta _{L} }{4\pi ^{2} } \cos 2\pi x
\end{equation}

\noindent
and

\begin{equation}\label{7}
\displaystyle u^{ScS} (x,x_{e} )=\frac{(x-x_{e} )^{2} }{2}
-\frac{\beta _{L} }{2\pi ^{2} } \left|\cos \pi x\right| .
\end{equation}

\begin{figure}[h!]
\centering
\includegraphics[width = 0.8\columnwidth]{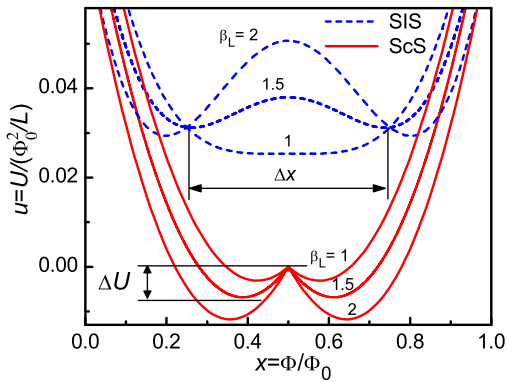}
\caption{\label{fig02}%
The potential energy of the ScS (solid line) and SIS (dashed line)
SQUIDs as a function of internal reduced magnetic flux for several
$\beta _{L} $ . The fixed external flux  $\Phi _{e} =\Phi _{0} /2$ .
The barrier height  $\Delta U$  and the inter-minima spacing $\Delta
x$  are shown in the curves for  $\beta _{L} =1.5$ .}
\end{figure}

The potential energy  $u^{SIS} (x,x_{e} )$  of the loop with a
tunnel junction has two or more local minima only at  $\beta _{L}
>1$. It can be symmetrized by applying fixed magnetic field  $\Phi
_{e} =\Phi _{0} /2\; \; (x_{e} =1/2)$ . Fig.\,\ref{fig01} shows
potential energy of the loops with SIS and ScS contacts as a
function of internal magnetic field at large  $\beta _{L} =12$  for
better visibility. The distinctive feature of the potential energy
$u^{ScS} (x,x_{e} )$  of an RF SQUID loop with the ScS contact is
the finite height of the "sharp" barrier between adjacent states for
any vanishingly small  $\beta _{L}$, i.e.  $L$  and/or  $I_{c} $ (at
zero temperature). Fig.\,\ref{fig02} displays potential energies of
the RF SQUIDs with SIS or ScS contact at several small $\beta _{L}$.
It is obvious from Fig.\,\ref{fig02} that the barrier separating the
two metastable current-carrying states vanishes at $\beta _{L} =1$
in the SIS SQUID while stays finite in the ScS SQUID.

The noise of thermal or any other origin causes switching between
the metastable states corresponding to the minima of  $U(\Phi )$ .
The switching rate for the white Gaussian noise with intensity
(variance)  $D$  for parabolic wells and  $\Delta U/D>>1$  is given
by the Kramers' formula \cite{19Kramers}:

\begin{equation}\label{8}
\displaystyle r_{K} =r_{0} \exp (-\Delta U/D) .
\end{equation}

In the case of thermal noise for  $T=0$ ,  $D=2k_{B} T$. In this
work we do not focus on any specific nature of the noise. The only
requirement imposed to ensure the adiabatic mode of the SQUID
switching is the limitation of the noise frequency band by a
frequency not exceeding the reciprocal relaxation time for the
internal flux in the SQUID loop  $1/\tau _{L} =R/L$. Although the
pre-exponential factor  $r_{0} $  is different for the "smooth"
potential of the SIS SQUID and "sharp" potential of the ScS SQUID
(details can be found in \cite{16Turutanov}), the exponent factor
dominates anyway. Adding a small periodic signal with frequency
$f_{s} $  to the external flux  $\Phi _{e} $  results in the
stochastic-resonance system motion in the noise-driven bistable
potential at

\begin{equation}\label{9}
\displaystyle r_{K} \approx 2f_{s} .
\end{equation}

Taking typical experimental figures  $L=3\cdot 10^{-10} \; H$,
$C=3\cdot 10^{-15} \; F$,  $R_{N} =1...10^{2} \; \Omega $,  $I_{c}
=10^{-5} ...10^{-6} \; A$, we found the McCamber parameter
describing the effect of capacitance  $\beta _{C} =2\pi R^{2} I_{c}
C/\Phi _{0} <1$. This case corresponds to the viscous,
non-oscillatory, motion and therefore the term with the second
derivative in Eq. (\ref{1}) can be neglected. Low signal frequency
$f_{s} =10\; Hz<<1/\tau _{L} $  and noise cut-off frequency  $f_{c}
\sim 10^{4} \; Hz<<1/\tau _{L} $  makes the problem adiabatic. This
allows us the time dependence of the external flux consider as the
time-dependent potential energy thus writing the motion equation in
the form

\begin{equation}\label{10}
\displaystyle \tau _{L} \frac{dx}{dt} +\frac{\partial
u(x,t)}{\partial x} =0 .
\end{equation}

With (\ref{6}) for the SIS junction, the Eq. (\ref{10}) reads as

\begin{equation}\label{11}
\displaystyle \frac{dx}{dt} =\frac{1}{\tau _{L} } [x_{e}
(t)-x+\frac{\beta _{L} }{4\pi } \sin 2\pi x] ,
\end{equation}

\noindent and with (\ref{7}) for the ScS contact, as

\begin{equation}\label{12}
\displaystyle \frac{dx}{dt} =\frac{1}{\tau _{L} } [x_{e}
(t)-x+\frac{\beta _{L} }{2\pi } (\sin \pi x) \,sgn\,(cos\pi x)].
\end{equation}

The external flux  $x_{e} (t)$  is the sum of the fixed bias flux
$x_{dc} =0.5$, the useful signal  $x_{ac} =a\sin 2\pi f_{s} t$  and
the noise flux  $x_{N}$. In the theory, the noise is often treated
as the  $\delta$-correlated white, Gaussian-distributed noise:
$x_{N} =\xi (t)$,  $\left\langle \xi (t)\cdot \xi
(t-t')\right\rangle =2D\delta (t-t')$. During the numerical
simulation, it is modeled by the quasi-random number generator
giving Gaussian-distributed values with variance  $D=\sigma ^{2} $
and the repetition period of about  $2^{90}$. The sampling rate is
$2^{16} $  when solving the equation by finite differences technique
that corresponds to the equivalent noise frequency band 32 kHz. So,
the noise can be considered as quasi-white for the stochastic
amplification of the signal with frequency  $f_{s} =10\; Hz$.

Eqs. (\ref{11}) and (\ref{12}) were solved by the Heun method
modified for the stochastic equations \cite{20Garcia}. Each run gave
an 8-s-long time series with its individual noise realization and
then were subjected to fast Fourier transform (FFT). The obtained
output spectral densities $S_{\Phi } (\omega )$  of the internal
flux were averaged over 30 runs. The spectral amplitude gain of the
weak periodic signal was determined as the ration of spectral
densities of the input (external) and output (internal) magnetic
fluxes:

\begin{equation}\label{13}
\displaystyle k(\omega )=S_{\Phi \; out}^{1/2} (\omega )/S_{\Phi \;
in}^{1/2} (\omega ) .
\end{equation}

\section{Results and discussion}

\begin{figure}[h!]
\centering
\includegraphics[width = 0.8\columnwidth]{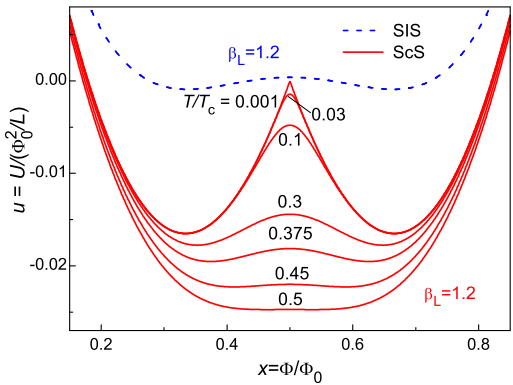}
\caption{\label{fig03}%
The potential energy of ScS SQUID as a function of the magnetic flux
in the loop at various reduced temperatures. Similar function for
the SIS SQUID at the same  $\beta _{L} $  is shown for comparison.}
\end{figure}

The current-phase relation for the ScS contact at a non-zero
temperature  $T$  \cite{14Beenakker,15Kulik}

\begin{equation}\label{14}
\displaystyle I_{s}^{ScS} (\varphi )=I_{c} \sin \frac{\varphi }{2}
\tanh \frac{\Delta (T)\cos \frac{\varphi }{2} }{2k_{B} T},
\end{equation}

\noindent where  $I_{s}^{ScS} (\varphi )$  is the contact
supercurrent, $I_{c} =\frac{\pi \Delta (T)}{eR_{N} } $  is the
contact critical current,  $\Delta (T)$  is the superconductor
energy gap (order parameter),  $\varphi $  is the order parameter
phase difference over the contact,  $k_{B} $  is the Boltzmann
constant,  $e$  is the electron charge,  $R_{N} $  is the contact
normal resistance, smoothes, tending, as approaching the
superconductor critical temperature  $T_{c}$, to the sine law that
describes the SIS junction. Consequently, the (reduced) potential
energy of the loop with ScS contact

\begin{equation}\label{15}
\displaystyle u_{T}^{ScS} =\frac{(x-x_{e} )^{2} }{2} -\frac{\beta
_{L} }{\pi ^{2} } t\ln \left(2\cosh \frac{\cos \pi x}{2t} \right) ,
\end{equation}

\noindent where  $t=\frac{T/T_{c} }{1.76\sqrt{1-(T/T_{c} )^{4} } }$,
also becomes closer in its shape to the SIS SQUID potential energy.
The singularity in the barrier top disappears, while the barrier
height decreases with the temperature rise. Fig.\,\ref{fig03}
illustrates the temperature evolution of ScS SQUID potential energy.
The SIS SQUID potential energy is also given in the Figure for the
comparison sake. In contrary to the zero-temperature case, the
potential barrier vanishes at a certain reduced temperature even for
$\beta _{L} >1$  (at  $T/T_{c} =0.5$  for the chosen  $\beta _{}
=1.2$) The inter-minima spacing decreases, too. The barrier heights
for the ScS- and SIS-SQUIDs become equal at a certain temperature
(at $T/T_{c} =0.375$  for the given case).

By substituting (\ref{15}) in(\ref{10}), we get the motion equation
for the reduced flux in ScS SQUID at finite temperatures $T$ ,
$0<T<T_{c} $ , similar to Eq. (\ref{11}) for the case  $T=0$. The
solution of this equation at various noise intensities with further
FFT as described above gives the spectral gain of a weak signal as a
function of the noise intensity, i.e. the SR curves.

\begin{figure}[h!]
\centering
\includegraphics[width = 0.8\columnwidth]{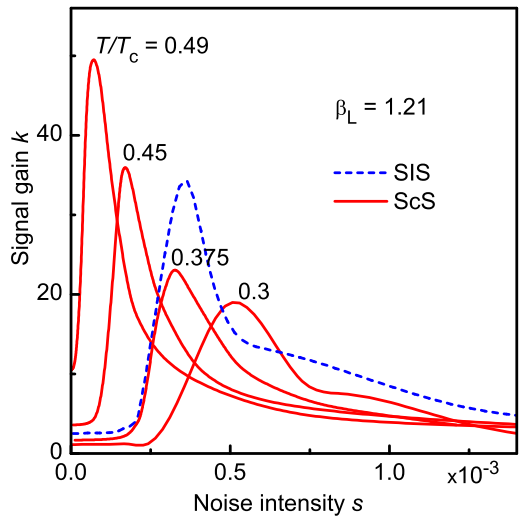}
\caption{\label{fig04}%
The weak harmonic signal gain as a function the noise intensity in
SIS and ScS SQUIDs at various reduced temperatures. The signal
amplitude is 0.001, frequency 10 Hz.}
\end{figure}

Fig.\,\ref{fig04} shows the SR curves in the ScS SQUID calculated
for several reduced temperatures and  $\beta _{L} =1.21$. Strictly
saying, $\beta _{L} $  parameter is indicated for  $T=0$, since it
renormalizes as the temperature rises. For the comparison, the SR
curve in a SIS SQUID loop is displayed in the figure for the same
$\beta _{L}$. One can see that, with the temperature rise, the
maximal gain in ScS SQUID increases due to lowering of the barrier.
This is much like a SIS SQUID behavior when  $\beta _{L} $  reduces
down to the unity \cite{16Turutanov}, so we consider the situation
as a "renormalization" of  $\beta _{L}$  in ScS SQUIDs.

When the barriers in the ScS and SIS SQUIDs are of the same height
($T/T_{c} =0.375$, see Fig.\,\ref{fig03}), the maximal gain is
observed at equal noise intensities for both SQUIDs while the
maximal gain values are different because of difference in the
potential shapes and inter-minima spacings  $\Delta x$
\cite{16Turutanov}.

According to the two-state theory for linear response
\cite{21McNamara} which does not accounts for inter-well dynamics,
the weak-signal gain is determined by the spacing  $\Delta x$
between the adjacent local minima of the system potential energy.
Taking the SR curves in Fig.\,\ref{fig04} for ScS and SIS SQUIDs
with almost equal maximal gain ( $T/T_{c} =0.45$  for ScS SQUID) and
looking up corresponding potential energies in Fig.\,\ref{fig04}, we
can see that  $\Delta x$ values are somewhat different. This implies
that the exact shapes of the potential profile should be involved in
accurate theoretical calculation. Nevertheless, it can be generally
stated that the stochastic amplification of an informational signal
in ScS SQUIDs with small $\beta _{L} $  will reveal its specific
features at ultra-low temperatures. The calculation shows the
potential barrier in ScS SQUID with  $\beta _{L} =0.1$  vanishes at
$T/T_{c} =0.045$ , that for niobium with its critical temperature
$T_{c} =9.25\; K$ corresponds to a millikelvin temperature range. At
such low temperatures, quantum effects will give a dominant
contribution in switching rate \cite{22Grifoni}. Moreover, the
"sharp" potential of a superconducting loop with ScS contact makes
quantum tunneling between separated wells much faster as compared to
a "common" SIS circuit. This feature allowed authors of
\cite{2Shnyrkov} to propose a superconducting qutrit-detector with
tunneling rate of 20-30 GHz (which was important there for large
energy level splitting). Thus, an interesting situation could emerge
in ScS SQUIDs at ultra-low temperatures when two characteristic time
scales would occur in the context of stochastic resonance, one for
classical hops over the barrier and other for quantum tunneling
through it.

\section{Conclusion}

The calculation of amplification of a weak harmonic signal on a
noisy background due to the stochastic resonance in an RF SQUID loop
with a clean ScS contact in the zero-temperature limit shows that
the SR in ScS SQUID is possible at any vanishingly small  $\beta
_{L} <1$  because of unusual "sharp" potential barrier between
metastable states whose height is always finite at  $T=0$.

With the ScS SQUID temperature rise, the top of the barrier
smoothes, its height gradually vanishes to zero, and the potential
of the ScS SQUID becomes much like that of "common" SIS SQUID. Thus,
the substantial difference in classic stochastic dynamics of
magnetic flux of ScS and SIS SQUIDs emerge at ultra-low
temperatures. The difference will become more distinctive if quantum
tunneling contribution is taken into account. These distinctions
should be considered, as useful or "parasitic" effect, in designing
deep-cooled devices involving superconducting loops with
incorporated Josephson junctions, such as SQUIDs and qubits.


\end{document}